\documentstyle[12pt]{article}
\begin{document}

\hfill DSF-T-6/94

\hfill hep-th/9404148

\renewcommand{\theequation}{\arabic{equation}}
\newcommand{\be}{\begin{equation}}
\newcommand{\ee}{\end{equation}}
\newcommand{\bea}{\begin{eqnarray}}
\newcommand{\eea}{\end{eqnarray}}
\newcommand{\bean}{\begin{eqnarray*}}
\newcommand{\eean}{\end{eqnarray*}}
\newcommand{\eqn}[1]{(\ref{#1})}
\renewcommand{\thefootnote}{\fnsymbol{footnote}}
\newcommand{\del}{\partial}

\def\up#1{\leavevmode \raise.16ex\hbox{#1}}

\def\tr{\mathop{\rm tr}\nolimits}
\def\vev#1{\left\langle #1\right\rangle}
\def\hc{{\rm h.c.}}

\newcommand{\npb}[3]{{\sl Nucl. Phys. }{\bf B#1} \up(19#2\up) #3}
\newcommand{\plb}[3]{{\sl Phys. Lett. }{\bf #1B} \up(19#2\up) #3}
\newcommand{\revmp}[3]{{\sl Rev. Mod. Phys. }{\bf #1} \up(19#2\up) #3}
\newcommand{\sovj}[3]{{\sl Sov. J. Nucl. Phys. }{\bf #1} \up(19#2\up) #3}
\newcommand{\jetp}[3]{{\sl Sov. Phys. JETP }{\bf #1} \up(19#2\up) #3}
\newcommand{\rmp}[3]{{\sl Rev. Mod. Phys. }{\bf #1} \up(19#2\up) #3}
\newcommand{\prd}[3]{{\sl Phys. Rev. }{\bf D#1} \up(19#2\up) #3}
\newcommand{\ijmpa}[3]{{\sl Int. J. Mod. Phys. }{\bf A#1} \up(19#2\up) #3}
\newcommand{\prl}[3]{{\sl Phys. Rev. Lett. }{\bf #1} \up(19#2\up) #3}
\newcommand{\physrep}[3]{{\sl Phys. Rep. }{\bf #1} \up(19#2\up) #3}
\newcommand{\journal}[4]{{\sl #1 }{\bf #2} \up(19#3\up) #4}

\setlength{\textheight}{9.0in}
\setlength{\textwidth}{5.75in}
\setlength{\topmargin}{-0.375in}
\hoffset=-.5in
\renewcommand{\baselinestretch}{1.17}
\setlength{\parskip}{6pt plus 2pt}


\newcounter{appendice}
\newcommand{\appendice}
{
\setcounter{equation}{0}
\renewcommand{\theequation}{\Alph{appendice}.\arabic{equation}}
\addtocounter{appendice}{1}
{\bf Appendix \Alph{appendice}}
}

\def\thebibliography#1{\section*{REFERENCES\markboth
 {REFERENCES}{REFERENCES}}\list
 {[\arabic{enumi}]}{\settowidth\labelwidth{[#1]}\leftmargin\labelwidth
 \advance\leftmargin\labelsep
 \usecounter{enumi}}
 \def\newblock{\hskip .11em plus .33em minus -.07em}
 \sloppy
 \sfcode`\.=1000\relax}
\let\endthebibliography=\endlist

\begin{center}
\vskip 2.5cm
{\LARGE \bf The Zero Tension Limit of the

Virasoro Algebra  and the Central

Extension}
\vskip 1.0cm
{\Large F.~LIZZI\footnote{
e-mail: LIZZI@NA.INFN.IT}}
\\
{\large  Dipartimento di Scienze Fisiche, Universit\`a di Napoli\\
and\\
I.N.F.N., Sez.\ di Napoli\\
Mostra d'Oltremare Pad.~19, 80125 Napoli.}
\vskip 0.4cm
\end{center}
\renewcommand{\baselinestretch}{1.17}

\begin{abstract}
We argue that the Virasoro algebra for the closed bosonic string can be
cast in a form which is suitable for the limit of vanishing string tension.
In this form the limit of the Virasoro algebra gives the null string
algebra. The anomalous central extension is seen to vanish as well
when $T\to 0$.
\end{abstract}

\vfill
\begin{center}
March 1994
\end{center}
\newpage
\setcounter{footnote}{1}

\newcommand{\pa}{\phi^\parallel}
\newcommand{\pe}{\phi^\perp}

A string theory is characterized by a dimensionful parameter, the string
tension, which plays the role played by the mass for point particle
theories. In this respect the zero tension limit of a string theory is in
some sense the analog of massless particles. The first to consider this
analogy was Schild \cite{Schild} (see also \cite{KarlhedeLindstrom}), which
considered strings whose geodesics are null surfaces. He called them `{\it
null strings}'. An action for null string was given in \cite{Baletal}, the
generalization to spinning and superstrings was started in
\cite{barcelosruiz} and an
Hamiltonian approach is due to Zheltukhin \cite{Zheltukhin}. The issues of
quantization where first discussed in \cite{lizzietal} and then in
\cite{grr1,grr2}. In \cite{lizzietal} was found that the quantization of
the null string does not give rise to critical dimensions, while in
\cite{grr1,grr2} was pointed out that the issue depends crucially on the
choice of ordering, and that ordering which gives the usual value of 26
for the number of critical dimensions are also consistent. When $T\to 0$
the Weyl invariance is substituted by conformal invariance (see
\cite{lindstrometal} and references therein), and the attempt to quantize
the theory keeping this invariance leads to restriction on the Hilbert
space. A very interesting connection with topological theories is made in
\cite{Isbergetal}.

In this short note our aim is more modest, we will not dwell upon the full
quantization, or the spectrum of the theory in the limit, more simply
we discuss the fate of critical dimensions, when
the zero tension limit of the usual Virasoro algebra is taken. Of course
the string tension does not appear as a parameter in the Virasoro algebra,
that can be taken to zero straightforwardly. What we will do is to show that
the algebra of constraints of the bosonic string can be expressed in such a
way that the tension does indeed appear as a parameter that can be sent to
zero, and that once this limit is taken, the resulting algebra is the one
of the null string. With this procedure the central extension can be seen
to go to zero as well when the limit is taken.

It is well known (see for example \cite{GreenSchwarzWitten}) that the
reparametrization algebra of the closed\footnote{In this paper we will
deal with closed bosonic strings, the generalizations being
straightforward.} bosonic string has two set of generators, called
$L_n$ and $\tilde L_n$ which satisfy the commutation algebra:
\be
[L_m,L_n,]=i(m-n)L_{m+n} +A_m \delta_{m+n,0} \label{Virasoro}
\ee
with an identical algebra for the $\tilde L$'s, which commute with the
$L$'s. Here $A_m={d\over 12}(m^3-m)$ is the so called central extension, a
consequence of the quantization, and which is the source of some of the
most interesting issues discussed in the past ten years or so.

Let us briefly trace the origins of \eqn{Virasoro}.
We start from the string coordinates $X^\mu(\sigma,\tau)$, a string action,
which after convenient gauge choices\footnote{For more details and
notations see \cite{GreenSchwarzWitten}.}, can be cast in the form:
\be
S=-{T\over 2} \int{d\sigma d\tau\sqrt{h} \eta^{\alpha\beta}\del_\alpha
X\cdot \del_\beta X } \label{action}
\ee
Where $\eta$ is the two dimensional Minkowski metric, and $T$ is a parameter
with the dimensions of inverse of the square of a length, the string
tension indeed.

The equations of motion of this action are of the wave equation form
\be
\left( {\del^2\over\del \sigma^2} - {\del^2\over\del \tau^2} \right) X^\mu =
0\ \ , \label{eqmot}
\ee
this means that the $X$, which are periodic in $\sigma$, can be expanded as
\be
X^\mu(\sigma,\tau)=X^\mu_R(\tau-\sigma)+X^\mu_L(\tau+\sigma)
\ee
with
\bea
X^\mu_R& =&{1\over 2} x_\mu +{\pi T\over 2}p^\mu(\tau-\sigma)
+\sqrt{\pi T} {i\over 2} \sum_{n\neq 0} {1\over n} \alpha^\mu_n
e^{-2in(\tau-\sigma)}
\nonumber\\
X^\mu_L& =&{1\over 2} x_\mu +{\pi T\over 2}p^\mu(\tau+\sigma)
+\sqrt{\pi T} {i\over 2} \sum_{n\neq 0} {1\over n} \tilde \alpha^\mu_n
e^{-2in(\tau+\sigma)} \label{fourier}
\eea

The residual gauge symmetry is represented by the constraints
\bea
\pe & = & P^2 +T^2 \left(\del_\sigma X\right)^2 = 0
\label{constraints1}\\
\pa & = & P\cdot X = 0 \label{constraints2}
\eea
Usually these constraints are split into left and right components and
become
\be
\dot X_R^2=\dot X_L^2=0 \ \ .
\ee

The Fourier modes of this last equation are the usual generators of the
Virasoro algebra:
\be
L_n={T\over 2} \int_0^\pi{e^{-2im\sigma} \dot X_R^2 d\sigma}
={1\over 2}\sum_{m=-\infty}^\infty \alpha_{n-m}\alpha_m
\ee

In the calculation of the algebra becomes crucial the issue of the ordering
of the $\alpha$'s. The issue is quickly resolved noticing that
\be
\alpha^\dagger_n=\alpha_{-n}
\ee
and noticing that the $\alpha$ satisfy harmonic oscillator--like equations.
This imposes the choice of normal ordering, with annihilator to the right of
creators, and consequently the appearance of the anomalous central
extension.

At the level of action obviously the $T\to 0$ limit cannot be taken in a
simple, we
will comment on the $T\to 0$ limit of the action later. The only place
where the limit can be safely made is at the level of the constraints
\eqn{constraints1} and \eqn{constraints2}.

In this limit the constraints become:
\bea
\hat \pe & = & P^2 = 0
\label{constnull1}\\
\hat \pa & = & P\cdot X = 0 \label{constnull2}
\eea
These are the constraints of the null string \cite{lizzietal} (here and in
the following we will indicate the quantities pertaining to the null string
with an hat). Let us very briefly see how one can find these constraints in
the context of null string, which we now very briefly describe.

As we mentioned earlier it is impossible to take the straightforward limit
of the action \eqn{action}, and an alternative must be found. Various
solution have been proposed, starting from the original `action principle'
described by Schild in his original article \cite{Schild}, other actions
have been described in
\cite{KarlhedeLindstrom,Baletal,HwangMarnelius,AmorBar}. For
example the action of \cite{Baletal} is given in terms of the quantity
$V^{\mu\nu}=\del_\sigma X^\mu\del_\tau X^\nu -
\del_\sigma X^\nu\del_\tau X^\mu $ which is some sort of generalization of
the velocity, and the auxiliary antisymmetric tensor  $\Sigma_{\mu\nu}$
constrained to satisfy $\Sigma_{\mu_1}^{\nu_1} \Sigma_{\mu_1}^{\nu_1} =
\ldots \Sigma_{\mu_1}^{\nu_1} \ldots \Sigma_{\mu_{d\over 2}}^{\nu_{d\over 2}} =
\varepsilon_{\mu_1\ldots\mu_d}\Sigma^{\mu_1\mu_2}\ldots
\Sigma{\mu_{d-1}\mu_d} =0$, with these fields the action
reads\footnote{For more details we refer to \cite{Baletal}.}:
\be
S_N=\int d\sigma d\tau=V^{\mu\nu}\Sigma^{\mu\nu}
\ee
After the elimination of $\Sigma$ the equations of motions are:
\be
\del_\tau^2 X=0\ \ .
\ee
They have to be contrasted with \eqn{eqmot}, where for the tensionful
string we had harmonic oscillators, for null strings we have free
particles. The constraints too are changed, and, as we mentioned above,
are given by \eqn{constnull1} and \eqn{constnull2}. The solution for the
equations of motion is now of the free particle kind:
\be
X(\sigma,\tau)=\hat x(\sigma) + \hat p(\sigma)\tau
\ee
and the Fourier expansions  which substitutes \eqn{fourier} are:
\bea
\hat x(\sigma) & =& a_ne^{in\sigma} \nonumber \\
\hat p(\sigma) & =& b_ne^{in\sigma}
\eea
The Fourier expansion for the constraints now gives:
\bea
\hat \pe_n & = &\sum_m a_{-m+n} a_m \nonumber \\
\hat \pa_n & = &\sum_m a_{-m+n} b_m
\eea
Considering the Poisson bracket between the $a$'s and the $b's$ which is
$\{a_n^\mu,b_m^\nu\}=\delta_{m,-n}^{\mu,\nu}$, one
finds the following (classical) algebra for the fourier components of the
constraints:
\bea
\{\hat \pe_n,\hat \pe_m\} & = & 0 \nonumber \\
\{\hat \pa_n,\hat \pe_m\} & = & (n-m)\hat \pa_{n+m} \nonumber \\
\{\hat \pa_n,\hat \pe_m\} & = & (n-m)\hat \pe_{n+m}
\eea

This is a classical result, to find the quantum one it is crucial the issue
of the normal ordering. It is in fact the commutation rules of the properly
ordered generators of the Virasoro Algebra to give the central extension.
In the null strings results are known to depend on the choice of ordering
\cite{lizzietal,grr1,grr2}, therefore we first choose the ordering (in the
tensionful case) to the usual normal ordering, which (remembering that
$[\alpha_{-n},\alpha_n]=1\ ,\alpha_n^\dagger=\alpha_{-n}$ reads:
\be
:\alpha_n \alpha_{-n}:= \alpha_{-n} \alpha_n \ \ .
\ee

We now have to calculate the algebra for $\pa$ and $\pe$ with this
ordering. The calculation could be very long, but is simplified greatly from
the
observation that
\bea
\pe&=&T(L_m+\tilde L_{-m})\nonumber\\
\pa&=&2(L_m-\tilde L_{-m})
\eea

Since we know the commutators of the $l$'s and $\tilde L$'s, including the
anomalous terms, we can just substitute them and calculate the commutators
of $\pa$ and $\pe$. A short calculation shows
\bea
{[\pa_m,\pa_{-m}]} &=& m \pa_0  \nonumber\\
{[\pe_m,\pe_{-m}]} &=& m \pa_0 T^2 \nonumber\\
{[\pe_m,\pa_{-m}]} &=& m \pe_0 +2 A_m T
\eea

It is now possible to take the $T\to 0$ limit.
The anomaly $A_m$ cancels in the first two cases, while it adds in the third.
This algebra is anomaly free only for the usual 26 dimensions and intercept.
For $T=0$ this algebra reduces to the one of null string \cite{lizzietal},
with no anomalous term which is proportional to the tension.

It is worth noticing at this point that the absence of the central
extension in the null string algebra depends on the choice of ordering. The
hermitian (or Weyl) ordering \cite{Zheltukhin,lizzietal} for which
$:x_{n} p_{n}:_{h.o.} = {1\over 2} (x_n p_n+p_n x_n)$ gives rise to no central
extension. If one starts from the null string, without any reference to
the tensionful string, then an alternative ordering \cite{grr1,grr2}, with
the positive modes of the $X$ and $P$ to the right of the negative ones,
is possible. In this case one finds the a central extension, together with
the value of 26 for the central extension. The choice of ordering of
course has consequences on the spectrum, which is continuous with the
hermitian ordering, massless with the ordering of \cite{grr1,grr2}. In
particular, we refer to \cite{grr2,lindstrometal} for discussions of the
issue of alternative orderings, and mass spectra.

The conclusion of this short note is that the $T\to 0$ limit of the string
can be taken at the level of the Virasoro algebra consistently, and that
the central extension (and of course with it the critical dimension)
vanishes with $T$. A result coinciding with the one of the null string,
with hermitian ordering of the operators.

I would like to thank I.~Giannakis for reawakening my interest in the zero
limit of the the string tension, thus convincing me of the opportunity to
publish this note.
\newpage
%
%
%

%
%
\end{document}